\documentstyle[preprint,aps,epsfig]{revtex}
\tightenlines
\newcommand{\beq}{\begin{equation}}
\newcommand{\eeq}{\end{equation}}
\newcommand{\ba}{\begin{eqnarray}}
\newcommand{\ea}{\end{eqnarray}}

\newcommand{\psibar}{\overline{\psi}}

\begin{document}
 
\draft
\title{
\begin{flushright}
\begin{minipage}{4 cm}
\small
hep-ph/0108179
\end{minipage}
\end{flushright}
Single transverse-spin asymmetry in Drell-Yan lepton angular distribution}

\author{Dani\"el Boer$^1$ and Jianwei Qiu$^2$}
\address{$^1$Dept.\ of Physics and Astronomy, Vrije Universiteit Amsterdam\\
             De Boelelaan 1081, 1081 HV Amsterdam, The Netherlands\\
         $^2$Department of Physics and Astronomy, Iowa State University \\
             Ames, Iowa, 50011, U.S.A.
}
\date{\today}
\maketitle

\begin{abstract}
We calculate a single transverse-spin asymmetry for the Drell-Yan
lepton-pair's angular distribution in perturbative QCD. At leading
order in the strong coupling constant, the asymmetry is expressed in
terms of a twist-3 quark-gluon correlation function
$T_F^{(V)}(x_1,x_2)$. 
In our calculation, the same result was obtained in both light-cone and
covariant gauge in QCD, while keeping explicit electromagnetic current
conservation for the virtual photon that decays into the lepton pair.
We also present a numerical estimate of the asymmetry and compare
the result to an existing other prediction. 
\end{abstract}
\vskip 0.1in

\section{Introduction}

It has been known for some time that cross sections involving 
a single transverse hadronic spin show significant asymmetries
relative to the spin direction \cite{Lambda-sts,pi-sts}.
Because of parity and time-reversal invariance, single longitudinal-spin
asymmetries vanish for most inclusive observables in high energy 
collisions; and single transverse-spin asymmetries open up 
a new domain of QCD dynamics: physics at twist-3, 
which is sensitive to the correlations between quark and gluon fields.  
With the active spin program at Brookhaven National Laboratory's 
relativistic heavy ion collider (RHIC), many single transverse-spin 
asymmetries have been proposed and could be observed in the near future 
\cite{QS-photon-sts,Boros,T-pi-sts,QS-pi-sts,HTS-dy-sts,BMT-dy-sts,BM-dy-sts}.

In this paper, we re-examine the single transverse-spin asymmetry 
in the Drell-Yan lepton angular distribution. Although the asymmetry 
was studied in Refs.~\cite{HTS-dy-sts,BMT-dy-sts,BM-dy-sts}, 
two different analytical expressions were derived. In Ref.\ \cite{BM-dy-sts} 
a formal argument, based on gauge vector independence, was given in favor of 
the result of Ref.\ \cite{BMT-dy-sts}, however, uncertainty about the result 
remained \cite{sts-wk-jp}. This motivates us to calculate the asymmetry in a
more transparent way, such that color and electromagnetic gauge invariance are
manifest. 

\begin{figure}
\begin{center}
\epsfig{figure=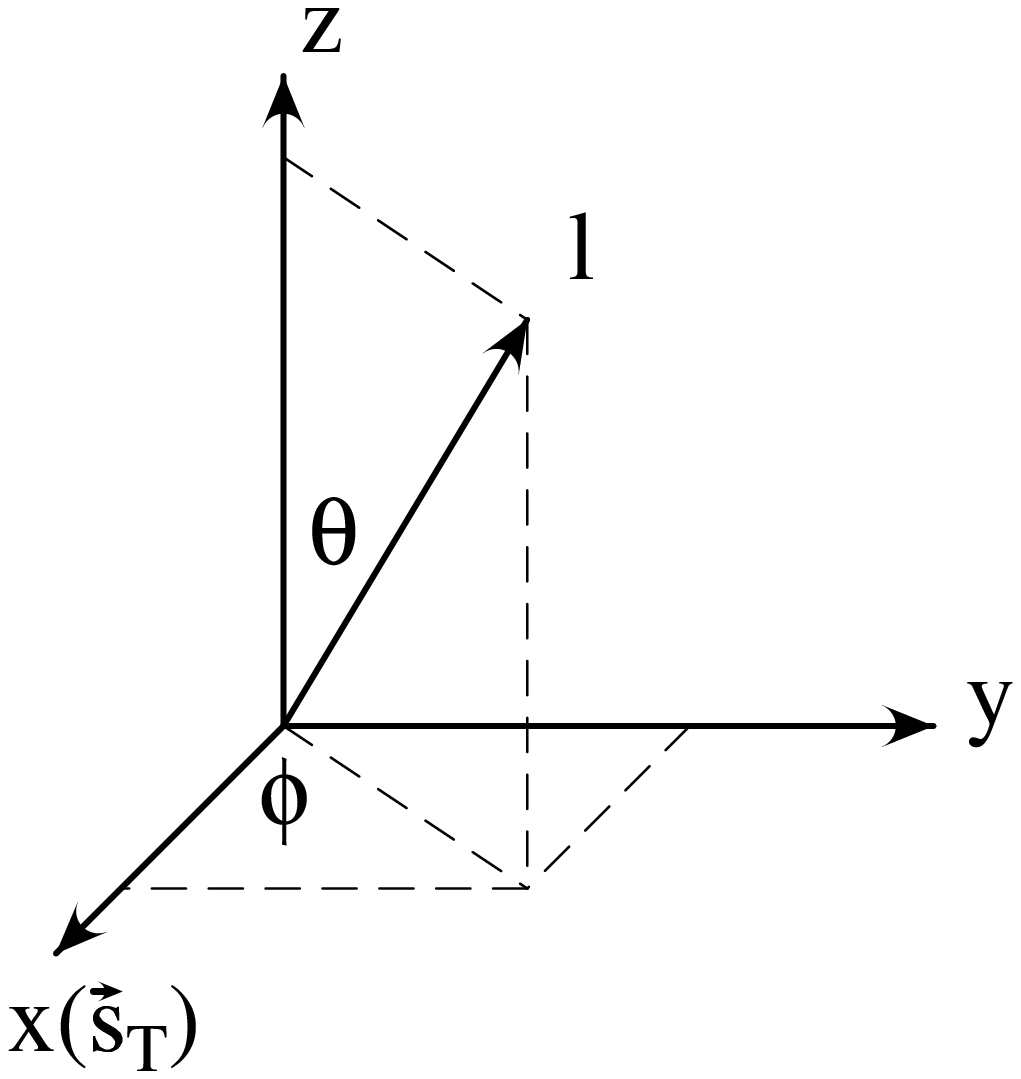,width=2.2in}
\end{center}
\caption{
Coordinate choice for the virtual photon's rest frame.
}
\label{fig1}
\end{figure}

For the polarized Drell-Yan process, $A(p,s_T)+B(p')\longrightarrow 
\gamma^*(Q)[\rightarrow \ell\bar{\ell}\;]+X$, we define the lepton-pair's
angular distribution in the virtual photon's rest frame ($\vec{Q}=0$). 
As shown in Fig.~\ref{fig1}, we choose the $z$-axis along the direction 
of polarized incoming hadron, $x$-axis along the direction of 
polarization vector $\vec{s}_T$, and the observed lepton momenta as 
\begin{eqnarray}
\ell^\mu &=& \frac{Q}{2} \left(
   1, \sin\theta\sin\phi, \sin\theta\cos\phi, \cos\theta \right);
\nonumber \\
\bar{\ell}^\mu &=& \frac{Q}{2} \left(
   1, -\sin\theta\sin\phi,-\sin\theta\cos\phi,-\cos\theta \right).
\label{l-def}
\end{eqnarray}
The $Q$ is the invariant mass of the virtual photon, and its 
four-momentum $Q^\mu = \ell^\mu + \bar{\ell}^\mu$.

The Drell-Yan single transverse-spin asymmetry is defined to be 
the difference of two spin-dependent cross sections, 
$d\sigma(s_T)/dQ^2d\Omega$ with opposite directions of polarization, 
divided by their sum,
\begin{eqnarray}
A_N &\equiv & \left.
\frac{1}{2} \left(
    \frac{d\sigma(s_T)}{dQ^2d\Omega} 
  - \frac{d\sigma(-s_T)}{dQ^2d\Omega} \right)
\right/
\frac{1}{2} \left(
    \frac{d\sigma(s_T)}{dQ^2d\Omega} 
  + \frac{d\sigma(-s_T)}{dQ^2d\Omega} \right)
\nonumber \\
&\equiv & \left.
    \frac{d\Delta\sigma(s_T)}{dQ^2d\Omega} 
\right/
    \frac{d\sigma}{dQ^2d\Omega} \, ,
\label{sts-def}
\end{eqnarray}
where the phase space $d\Omega = d\cos\theta d\phi$.

In Ref.~\cite{HTS-dy-sts}, Hammon, Teryaev, and Sch\"{a}fer (HTS) 
calculated the Drell-Yan single transverse-spin asymmetry at tree-level 
in perturbative QCD with the light-cone gauge, and obtained 
\begin{equation}
A_N^{(HTS)}= g \left[ 
               \frac{\sin 2\theta \sin \phi}{1+ \cos^2 \theta} \right]
           \frac{1}{Q} \,
           \frac{\left[ T(x,x) -x \; \frac{dT(x,x)}{dx}\right]}
                {q(x)}\, ,
\label{HammonA}
\end{equation}
where $g=\sqrt{4\pi\alpha_s}$ is the coupling constant for strong 
interaction, and the summation over quark flavor was suppressed.    
The $q(x)$ is the normal unpolarized quark distribution.  The 
$T(x,x)=-T_F^{(V)}(x,x)$ was called a twist-3 soft gluon pole function, 
with the $T_F^{(V)}(x_1,x_2)$ defined as \cite{QS-photon-sts}
\begin{eqnarray}
T_{F}^{(V)}(x_1,x_2) &=& 
\int \frac{dy_1^- \; dy_2^-}{4\pi} e^{i x_1 p^+ y_1^-
+ i (x_2-x_1) p^+ y_2^-} 
\nonumber \\
&\times &
\langle p,\vec{s}_T | \psibar(0) \, \gamma^+ \,  
\left[ \epsilon^{s_T \sigma n \bar{n}} F_\sigma{}^+(y_2^-) \right]\,
\psi(y_1^-) | p,\vec{s}_T \rangle\, .
\label{TFV-def}
\end{eqnarray}
The $\pm$ components of any vector $k$ are defined in terms of two 
lightlike vectors $n$ and $\bar{n}$ as $k^+=k\cdot n$
and $k^-=k\cdot \bar{n}$.  These lightlike vectors are chosen 
such that $\bar{n}\cdot n=1$.  In our choice of frame, the 
momentum of the polarized hadron $p$ is up to a mass term 
proportional to $\bar{n}$.

In Ref.~\cite{BMT-dy-sts}, Boer, Mulders, and Teryaev (BMT) also
investigated the Drell-Yan single transverse-spin asymmetry.  
A similar but different
result was derived. This result was re-expressed in the notation of 
Ref.~\cite{HTS-dy-sts} and was found to be \cite{BM-dy-sts} 
\begin{equation}
A_N^{(BMT)} = g \left[ 
               \frac{\sin 2\theta \sin \phi}{1+ \cos^2 \theta} \right]
           \frac{1}{Q} \,
           \frac{T(x,x)}{q(x)}\, ,
\label{BoerA}
\end{equation}
which differs from $A_N^{(HTS)}$ in Eq.~(\ref{HammonA}) by the term 
proportional to $x \; dT(x,x)/dx$.   

In this paper, we calculate the leading order contributions to
the Drell-Yan single transverse-spin asymmetry in terms of a 
generalized QCD factorization theorem \cite{QS-photon-sts,QS-fac-91}.  
In order to test the gauge invariance of our result, we derive the
asymmetry in light-cone gauge as well as in covariant gauge of QCD 
while keeping the electromagnetic current conservation of 
the virtual photon explicit. The latter also makes the calculation much more
transparent than the one of Ref.\ \cite{BMT-dy-sts}, which uses so-called 
effective T-odd distribution functions. 
In our present calculation we find that the results are independent of the 
gauge choices and we confirm the result of Ref.\ \cite{BMT-dy-sts,BM-dy-sts}.

The derivative term in HTS's result in Eq.~(\ref{HammonA}) 
has been found in single transverse-spin 
asymmetries in other processes, like prompt photon production 
\cite{QS-photon-sts} and pion production in hadron-hadron scattering 
\cite{QS-pi-sts}, where the derivative term arises from collinear
expansions of the partonic parts while calculations were performed in
covariant gauge. However, we argue below that 
in terms of the generalized factorization theorem \cite{QS-fac-91}, 
there is no derivative term for the Drell-Yan asymmetry 
at the tree-level while it can appear in higher order corrections.

The rest of this paper is organized as follows. In the next section, we
derive the single transverse-spin asymmetry in the Drell-Yan lepton angular
distribution in perturbative QCD with a choice of light-cone gauge.
In our calculation, we monitor the electromagnetic current conservation of 
the virtual photon by calculating the strong interaction part and the
leptonic part separately. In addition, we show explicitly why there is no 
derivative term. As a cross check, we use
the method developed in Ref.~\cite{QS-photon-sts} to calculate the same
Drell-Yan single transverse-spin asymmetry in a covariant gauge
in Sec.~III. An identical result is obtained.  
Finally, we present a numerical estimate of the
asymmetry and our conclusions in Sec.~IV.

\section{Asymmetry calculated in QCD light-cone gauge}

In order to calculate the Drell-Yan single transverse-spin asymmetry, 
$A_N$ in Eq.~(\ref{sts-def}), we need to calculate both the  
unpolarized cross section, $d\sigma/dQ^2d\Omega$, and the
polarized cross section, $d\Delta\sigma(s_T)/dQ^2d\Omega$.

The unpolarized Drell-Yan cross section can be factorized 
according to the QCD factorization theorem \cite{CSS-fac,Bodwin-fac}, 
\begin{equation}
\frac{d\sigma_{AB\rightarrow\ell\bar{\ell}(Q)X}}{dQ^2d\Omega}
=\sum_{a,b} \int dx \phi_{a/A}(x) \, dx' \phi_{b/B}(x')\,
  \frac{d\hat{\sigma}_{ab\rightarrow\ell\bar{\ell}(Q)X}}
       {dQ^2d\Omega}
\label{DY-LO-fac}
\end{equation}
where $\sum_{a,b}$ run over all possible parton flavors, 
the $\phi$'s are normal parton distributions, and 
$d\hat{\sigma}/dQ^2d\Omega$ are perturbatively calculable partonic hard 
parts.  At the lowest order, the partonic hard part is given 
by the $q\bar{q}\rightarrow \gamma^*(Q) \rightarrow \ell\bar{\ell}$\ 
tree-level diagram, and can be expressed as
\begin{equation}
\frac{d\hat{\sigma}_{q\bar{q}\rightarrow\ell\bar{\ell}(Q)X}}
     {dQ^2d\Omega}
= e_q^2 \left(\frac{\alpha_{em}^2}{4\hat{s}}\right)
  H^{\mu\nu}(xp,x'p')\, L_{\mu\nu}(\ell,Q-\ell) 
  \left(\frac{1}{Q^2}\right)^2 
  \delta(Q^2-(xp+x'p')^2)\, ,
\label{DY-LO-p}
\end{equation}
where $\hat{s}=(xp+x'p')^2\approx xx'S$ with total c.m.\ energy 
$S=2p\cdot p'$.
The parton level hadronic tensor $H^{\mu\nu}$ is given by
\begin{equation}
H^{\mu\nu}(xp,x'p')= \left(\frac{1}{3}\right) \big[
(xp)^\mu (x'p')^\nu + (xp)^\nu (x'p')^\mu 
- 2(xp)\cdot(x'p') g^{\mu\nu} \big]
\label{DY-H-LO}
\end{equation}
with the explicit color factor (1/3).  
The corresponding leptonic tensor is 
\begin{equation}
L_{\mu\nu}(\ell,Q-\ell) = 4 \left[
\ell_\mu (Q-\ell)_\nu + (Q-\ell)_\mu \ell_\nu 
-\ell\cdot Q\ g_{\mu\nu} \right]\, .
\label{DY-L-LO}
\end{equation}
Both hadronic and leptonic tensors show explicit electromagnetic 
current conservation: $Q_{\mu}H^{\mu\nu}=0$ and 
$Q^\mu L_{\mu\nu}=0$, which will be an issue below 
when we calculate the twist-3 contributions to the asymmetry.

By contracting the hadronic and leptonic tensors, we obtain 
the lowest order unpolarized Drell-Yan cross section in the virtual
photon's rest frame,
\begin{equation}
\frac{d\sigma_{AB\rightarrow\ell\bar{\ell}(Q)X}}{dQ^2d\Omega}
= \frac{4\pi\alpha_{em}^2}{9Q^2} \sum_{q} e_q^2
\int dx\, q(x) \, dx'\, \bar{q}(x')\,
\left[\frac{1}{4\pi}\left(\frac{3}{4}\right)
   \left[1+\cos^2\theta\right] \delta(Q^2-xx'S)\right] ,
\label{DY-LO}
\end{equation}
where $\sum_q$ runs over all quark and antiquark flavors and 
the factorization scale dependence of the parton distributions 
is suppressed.

\begin{figure}
\begin{center}
\epsfig{figure=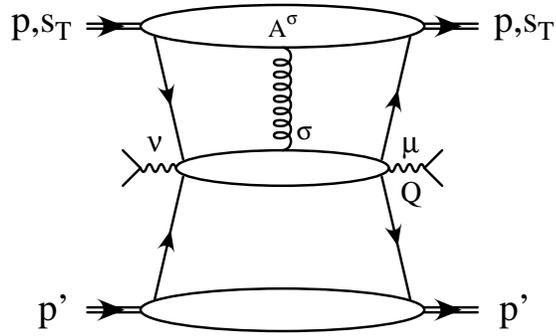,width=3.0in}
\end{center}
\caption{
General diagram contributing to the Drell-Yan single 
transverse-spin asymmetry.
}
\label{fig2}
\end{figure}

\begin{figure}
\begin{center}
\epsfig{figure=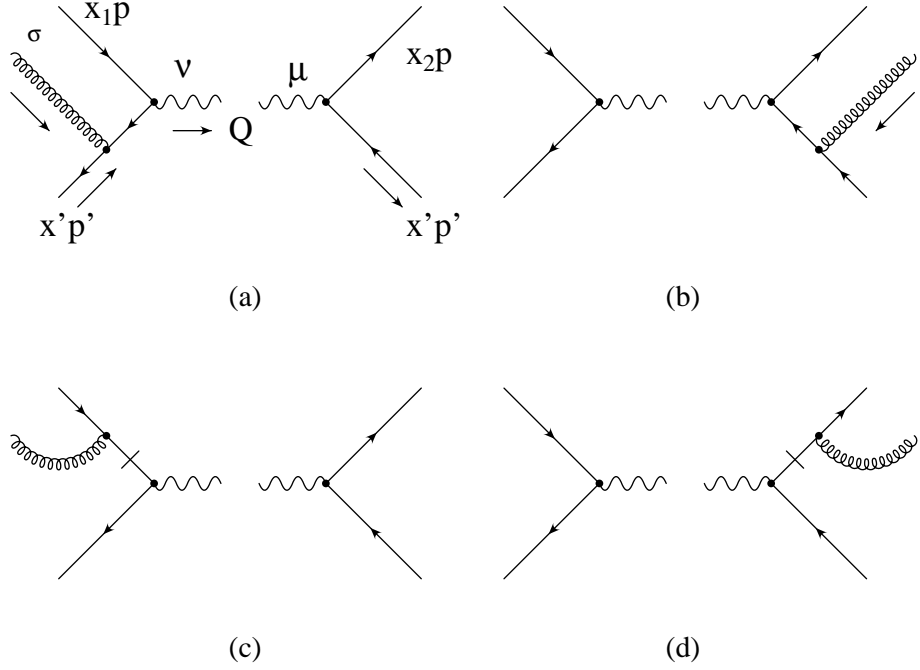,width=4.8in}
\end{center}
\caption{
Lowest order Feynman diagrams contributing to the
Drell-Yan single transverse-spin asymmetry.
}
\label{fig3}
\end{figure}

The single transverse-spin asymmetry is a twist-3 effect.  In contrast to 
Eq.~(\ref{DY-LO-fac}), the Drell-Yan single transverse-spin asymmetry
has the following generalized factorization formula 
\cite{QS-photon-sts} 
\begin{equation}
\frac{d\Delta\sigma_{AB\rightarrow\ell\bar{\ell}(Q)X}(s_T)}
     {dQ^2d\Omega}
= \sum_{a,b} \int dx_1 dx_2\, T_{a/A}^{(3)}(x_1,x_2;s_T) 
             \int dx' \phi_{b/B}(x')\ 
             \frac{1}{Q}\
\frac{d\Delta\hat{\sigma}_{ab\rightarrow\ell\bar{\ell}(Q)X}(s_T)}
     {dQ^2d\Omega} ,
\label{DY-sts-fac}
\end{equation}
where $T_{a/A}^{(3)}(x_1,x_2;s_T)$ are twist-3 three-parton 
correlation functions with a dimension of mass and 
$d\Delta\hat{\sigma}(s_T)/dQ^2d\Omega$ are perturbatively calculable 
partonic parts.  The explicit $1/Q$ factor in Eq.~(\ref{DY-sts-fac})
is a signal of a twist-3 effect and takes care of the 
mass dimension of the twist-3 correlation functions.  
In order to extract the lowest order contribution to the Drell-Yan single 
transverse-spin asymmetry, 
we need to start with lowest order diagrams 
with two-fermion and one gluon lines from the polarized hadron, 
as shown in Fig.~\ref{fig2} \cite{QS-photon-sts,QS-fac-91,Qiu-T4}.  
Because of the virtual photon of the Drell-Yan process, 
contributions from Fig.~\ref{fig2} can be written in general as
\begin{equation}
I\equiv \left(\frac{1}{Q^2}\right)^2
        W^{\mu\nu}(Q)\ L_{\mu\nu}(\ell,Q-\ell) 
\label{DY-W-L}
\end{equation}
with the leptonic tensor $L_{\mu\nu}$ given in Eq.~(\ref{DY-L-LO}).
After collinear expansion of the parton's momenta, and 
separation of fermion traces \cite{QS-photon-sts,QS-fac-91,Qiu-T4}, 
the leading contributions to the hadronic tensor $W^{\mu\nu}$ 
from the diagram in Fig.~\ref{fig2} can be 
expressed in the following factorized form,
\begin{equation}
W^{\mu\nu}(Q) = \sum_{q} e_q^2
   \int dx_1 dx_2\, T_{q}^{\sigma}(x_1,x_2;s_T) 
   \int dx'\, \bar{q}(x')\ 
   H^{\mu\nu}_{\sigma}(x_1,x_2,x';Q)\, ,
\label{W-sts-fac}
\end{equation}
where $\sum_q$ runs over all quark and antiquark flavors.  
The hadronic matrix element $T_{q}^{\sigma}(x_1,x_2;s_T)$ is 
given by
\begin{eqnarray}
T_q^{\sigma}(x_1,x_2;s_T) &=&
\int \frac{p^+dy_1^-}{2\pi}\, \frac{p^+dy_2^-}{2\pi} \
     {\rm e}^{i x_1 p^+ y_1^-}\, {\rm e}^{i (x_2-x_1) p^+ y_2^-} 
\nonumber \\
&\times &
\langle p,\vec{s}_T | \psibar_q(0) 
  \left(\frac{\gamma^+}{2p^+}\right)  
  A^\sigma(y_2^-) \, \psi_q(y_1^-) | p,\vec{s}_T \rangle\, .
\label{T-sigma-def}
\end{eqnarray}
where $A^\sigma$ is the gluon field.  The partonic part
$H^{\mu\nu}_{\sigma}$ in Eq.~(\ref{W-sts-fac}) is given by
the four Feynman diagrams in Fig.~\ref{fig3} with 
the quark lines contracted by $\frac{1}{2}\gamma\cdot p$, 
and antiquark lines contracted by $\frac{1}{2}\gamma\cdot p'$
\cite{QS-photon-sts,QS-fac-91,Qiu-T4}. 
As we will demonstrate below, the inclusion of all four 
diagrams are very important for preserving electromagnetic
current conservation.  The diagrams \ref{fig3}(c) and 
\ref{fig3}(d) play the same role as the twist-3 contribution of 
diagram Fig.~1 of Ref.\ \cite{BMT-ZeuthenProc} in the calculation 
of Ref.~\cite{BMT-dy-sts}.  

It is important to note that the $T_q^{\sigma}(x_1,x_2;s_T)$ 
in Eq.~(\ref{T-sigma-def}) is not necessarily a twist-3 matrix 
element, and the hadronic tensor $W^{\mu\nu}$ in 
Eq.~(\ref{W-sts-fac}) could have twist-2, twist-3, and even 
higher twist contributions \cite{Qiu-T4}.  In the rest of this 
section, we show how to extract the twist-3 contributions to 
the single transverse-spin asymmetry from the $W^{\mu\nu}$ 
in Eq.~(\ref{W-sts-fac}) in light-cone gauge.  
We will present a similar derivation in a 
covariant gauge in the next section.

In $n\cdot A(y_2^-)=0$ light-cone gauge, the  
leading contribution of the gluon field 
$A^\sigma$ comes from its transverse components 
(i.e., $\sigma = 1,2$) \cite{LQS-DIS}, and the 
corresponding gluon field strength is given by 
\begin{equation}
F^{+\sigma}(y^-_2) = n^\rho \partial_\rho\, A^\sigma(y_2^-)\, .
\label{F-LC}
\end{equation}
Therefore, in the light-cone gauge, 
we need to make the following replacement,
\begin{equation}
A^\sigma(y_2^-) \longrightarrow \frac{i}{(x_2-x_1)p^+}\,
 F^{+\sigma}(y_2^-)
\label{A-2-F-LC}
\end{equation}
in the matrix element $T^\sigma(x_1,x_2;s_T)$ 
to convert the gluon field $A^\sigma(y_2^-)$ into the gluon field 
strength $F^{+\sigma}(y_2^-)$, the same as that 
in the twist-3 matrix element $T_F^{(V)}(x_1,x_2)$ in 
Eq.~(\ref{TFV-def}).

Because of different behavior of the fermion propagators, 
we derive the partonic tensor 
$H^{\mu\nu}_{\sigma}$ in Eq.~(\ref{W-sts-fac}) in two steps.  
We first calculate the contributions from the two diagrams 
in Fig.~\ref{fig3}(a) and \ref{fig3}(b). With the 
Lorentz index $\sigma$ in the transverse direction, we obtain
\begin{equation}
H^{\mu\nu}_{\sigma}(a+b) = 
g\left( \frac{x_2-x_1}{x_2-x_1+i\epsilon} \right)
\left[ g^{\mu}_\sigma \left(\frac{p}{x'}\right)^\nu
       \delta(Q^2-x_2x'S)
     + g^{\nu}_\sigma \left(\frac{p}{x'}\right)^\mu
       \delta(Q^2-x_1x'S) \right].
\label{H-LC-ab}
\end{equation}
Clearly, the partonic part from the two diagrams violates 
electromagnetic current conservation.  

After the collinear expansion, the fermion propagators in 
the other two diagrams in Fig.~\ref{fig3} are on mass-shell 
and contributions from these two diagrams are divergent.  
As shown in Refs.~\cite{QS-fac-91,Qiu-T4}, the divergent 
part of these propagators corresponds to the long-distance 
contributions that should be included into the twist-2 
quark distribution;
and these propagators also have a finite contact 
contribution, which is one twist higher. The concept of special
propagator was introduced in Ref.~\cite{Qiu-T4} to identify 
the high twist piece of each propagator.  
Graphically, the special propagator is represented 
by a normal propagator line plus a bar as shown 
in Fig.~\ref{fig3}(c) and \ref{fig3}(d).  The Feynman rule 
for a special propagator depends on the gauge choice 
because of the nature of high twist contributions. In 
light-cone gauge, for example, the rule for a special 
quark propagator, merged from incoming quark and gluon lines 
in Fig.~\ref{fig3}(c), is given by \cite{Qiu-T4} 
\begin{equation}
\frac{i\gamma\cdot (x_2p)}{(x_2p)^2+i\epsilon}
\longrightarrow 
\frac{i\gamma\cdot n}{2x_2 p\cdot n}
\left[ \frac{x_2-x_1}{x_2-x_1+i\epsilon} \right]
\label{SP-q-LC}
\end{equation}
where the ``$i\epsilon$'' can be either derived following the 
example given in the next section, or obtained directly from 
the prescriptions given in the Appendix of 
Ref.~\cite{MQ-Recom-86}.  With the 
special quark propagator in Eq.~(\ref{SP-q-LC}), we derive 
the twist-3 contributions from the diagrams in Fig.~\ref{fig3}(c) 
and \ref{fig3}(d),
\begin{equation}
H^{\mu\nu}_{\sigma}(c+d) = 
g\left( \frac{x_2-x_1}{x_2-x_1+i\epsilon} \right)
\left[ g^{\mu}_\sigma \left(-\frac{p'}{x_2}\right)^\nu
       \delta(Q^2-x_2x'S)
     + g^{\nu}_\sigma \left(-\frac{p'}{x_1}\right)^\mu
       \delta(Q^2-x_1x'S) \right].
\label{H-LC-cd}
\end{equation}
Combining the short-distance twist-3 contributions 
from all four diagrams in Fig.~\ref{fig3}, we obtain
\begin{eqnarray}
H^{\mu\nu}_{\sigma} &=& 
g\left( \frac{x_2-x_1}{x_2-x_1+i\epsilon} \right)
\bigg[ g^{\mu}_\sigma 
       \left(\frac{p}{x'}-\frac{p'}{x_2}\right)^\nu
       \delta(Q^2-x_2x'S)
\nonumber \\
&\ & {\hskip 1.3in}     
     + g^{\nu}_\sigma 
       \left(\frac{p}{x'}-\frac{p'}{x_1}\right)^\mu
       \delta(Q^2-x_1x'S) \bigg].
\label{H-LC}
\end{eqnarray}
The partonic part $H^{\mu\nu}_{\sigma}$ 
with complete twist-3 contributions from all four diagrams 
in Fig.~\ref{fig3} clearly respects the electromagnetic 
current conservation.

Having the partonic part $H^{\mu\nu}_{\sigma}$ 
in Eq.~(\ref{H-LC}) and the matrix element $T^\sigma$ 
in Eq.~(\ref{T-sigma-def}) with its gluon field 
replaced by the corresponding field strength as 
in Eq.~(\ref{A-2-F-LC}), we derive the twist-3 contributions 
to the hadronic tensor $W^{\mu\nu}(Q)$ in Eq.~(\ref{W-sts-fac}),
\begin{eqnarray}
W^{\mu\nu}(Q) &=& \sum_{q} e_q^2
   \int dx\, T_{F_q}^{(V)}(x,x) 
   \int dx'\, \bar{q}(x') \left(
   \delta(Q^2-xx'S)\, \frac{g}{2}\, \epsilon^{s_T\sigma n \bar{n}}\right)
\nonumber \\
&\ & {\hskip 0.4in} \times
\left[ g^{\mu}_\sigma 
       \left(\frac{p}{x'}-\frac{p'}{x}\right)^\nu
     + g^{\nu}_\sigma 
       \left(\frac{p}{x'}-\frac{p'}{x}\right)^\mu
\right].
\label{W-sts-LO}
\end{eqnarray}
In deriving the above result, we took the imaginary part of the single 
pole, 
\begin{equation}
\frac{1}{x_2-x_1+i\epsilon} \longrightarrow
-\pi\, i\, \delta(x_2-x_1)\, ,
\label{IM-pole}
\end{equation}
such that the $i$ from the imaginary part cancels the $i$ 
in Eq.~(\ref{A-2-F-LC}) to ensure a real contribution 
to the single transverse-spin asymmetry \cite{QS-photon-sts}.  
The apparent factor $x_2-x_1$ in numerator of Eq.~(\ref{H-LC}) 
cancels the $1/(x_2-x_1)$ in Eq.~(\ref{A-2-F-LC}), 
which is the result of converting the matrix element $T^\sigma$ 
to the twist-3 correlation function $T_F^{(V)}$.  
It is clear that the leading order twist-3 hadronic tensor 
$W^{\mu\nu}(Q)$ in Eq.~(\ref{W-sts-LO}) respects electromagnetic 
current conservation,
\begin{equation}
 W^{\mu\nu}(Q)\, Q_\mu 
= W^{\mu\nu}(Q)\, Q_\nu = 0\, ,
\label{EM-cons}
\end{equation}
with the virtual photon's momentum $Q^\mu=(xp)^\mu+(x'p')^\mu$.

By contracting the hadronic tensor $W^{\mu\nu}(Q)$ with the 
leptonic tensor $L_{\mu\nu}$ in Eq.~(\ref{DY-L-LO}), and 
using Eq.~(\ref{DY-W-L}), we derive  
the leading order twist-3 contributions from the general diagram 
in Fig.~\ref{fig2}, 
\begin{equation}
I^{(3)} = \left(\frac{1}{Q^2}\right)^2 
   \sum_{q} e_q^2
   \int dx\, T_{F_q}^{(V)}(x,x) 
   \int dx'\, \bar{q}(x') \left[
     -\, \frac{g}{Q}\; SQ^2 (\sin 2\theta\sin\phi)
     \delta(Q^2-xx'S) \right],
\label{WL-3-LO}
\end{equation}
where $\sum_q$ runs over all quark and antiquark flavors.
Multiplying the twist-3 contributions in
Eq.~(\ref{WL-3-LO}) by the overall flux, coupling, 
and phase space factor, 
$(\alpha_{em}^2/4S)=1/(2S)\, e^4\, (1/32\pi^2)$, as well as 
a color factor $(1/3)$, we obtain the leading order 
polarized cross section, 
\begin{eqnarray}
\frac{d\Delta\sigma_{AB\rightarrow\ell\bar{\ell}(Q)X}(s_T)}
     {dQ^2d\Omega}
&=& 
\frac{4\pi\alpha_{em}^2}{9Q^2} \sum_{q} e_q^2
\int dx\, T_{F_q}^{(V)}(x,x) \, dx'\, \bar{q}(x')
\nonumber \\
&\ & \times
\left[\frac{1}{4\pi}\left(\frac{3}{4}\right)
   \left[-\frac{g}{Q}\,\sin2\theta\sin\phi\right] 
   \delta(Q^2-xx'S)\right] .
\label{DY-sts-LO}
\end{eqnarray}
Use $T_q(x,x)=-T_{F_q}^{(V)}(x,x)$, and divide the polarized 
cross section in Eq.(\ref{DY-sts-LO}) 
by the unpolarized cross section in Eq.~(\ref{DY-LO}), 
we derive our result for the single transverse-spin asymmetry 
for Drell-Yan lepton angular distribution, 
\begin{equation}
A_N = \sqrt{4\pi\alpha_s}\, \left[ 
      \frac{\sin 2\theta \sin \phi}{1+ \cos^2 \theta} \right]
      \frac{1}{Q} \,
      \frac{\sum_q e_q^2 \int dx\, T_q(x,x)\, \bar{q}(Q^2/xS)}
           {\sum_q e_q^2 \int dx\, q(x)\, \bar{q}(Q^2/xS)},
\label{BQ-AN-Full}
\end{equation}
where $\sum_q$ runs over all quark and antiquark flavors.
If we suppress the sum over quark flavors, as what was done in 
Refs.~\cite{HTS-dy-sts,BMT-dy-sts}, the asymmetry $A_N$ in 
Eq.~(\ref{BQ-AN-Full}) is reduced to the $A_N^{(BMT)}$ 
in Eq.~(\ref{BoerA}). 

\begin{figure}
\begin{center}
\epsfig{figure=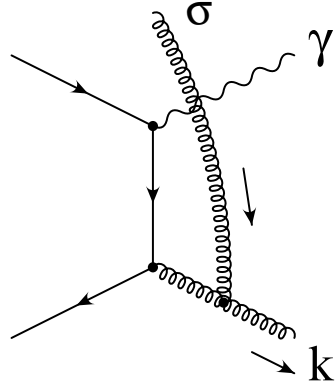,width=1.8in}
\end{center}
\caption{
Sample Feynman diagram generating a double soft-gluon pole.
}
\label{fig4}
\end{figure}

With three partons from the polarized hadron, the general Feynman 
diagram in Fig.~\ref{fig2} can have both soft-gluon as well as 
soft-quark poles \cite{QS-photon-sts}.  Because of the infrared 
behavior of soft gluons, the soft-gluon poles from the general 
diagram in Fig.~\ref{fig2} can be either a 
single pole, like the $1/(x_2-x_1+i\epsilon)$ in 
Eq.~(\ref{H-LC}), or a double pole, such as 
$1/(x_2-x_1+i\epsilon)^2$.  The single pole leads to a dependence 
on $T(x,x)$ while the double pole results into the derivative term
$xdT(x,x)/dx$.  For example, consider a diagram in 
Fig.~\ref{fig4}, which gives a higher order correction to the Drell-Yan
single transverse-spin asymmetry if the photon is virtual.  
If the photon is real, it gives a leading order contribution to the 
single transverse-spin asymmetry in prompt photon production
\cite{QS-photon-sts}. 
In light-cone gauge, this diagram potentially has  
a double soft-gluon pole: one from the gluon propagator and 
the other from the gluon attached to the polarized hadron.  
However, in general, one of the two potential poles 
might be canceled by a vanishing numerator.  
But, if the outgoing gluon in Fig.~\ref{fig4} 
has nonvanish transverse momentum, both soft poles could survive 
and lead to a double soft-gluon pole \cite{QS-photon-sts}.  

However, the potential double poles never survive at the lowest 
order single transverse-spin asymmetry in the Drell-Yan angular 
distribution calculated in this paper.  Consider the diagram in 
Fig.~\ref{fig3}(a), we have potential poles from the quark 
propagator, 
$$
i\; \frac{(x_1-x_2) \gamma\cdot p -x'\gamma\cdot p'}
         {-(x_1-x_2)x'S+i\epsilon}\, ,
$$
and the $i/(x_2-x_1)p^+$ in Eq.~(\ref{A-2-F-LC}) from the gluon 
attached to the polarized hadron in light-cone gauge.
In terms of the generalized factorization theorem 
\cite{QS-photon-sts,QS-fac-91}, momenta of all partons 
entering the partonic short-distance hard parts 
have only components collinear to their respective hadron momenta.
For the leading order Drell-Yan process, 
the incoming quark and antiquark lines in Fig.~\ref{fig3} 
have to be contracted with $\frac{1}{2}\gamma\cdot p$ and 
$\frac{1}{2}\gamma\cdot p'$ to ensure the twist-3 contributions
and color gauge invariance at this twist.  
Any other combination will result into power suppressed 
corrections \cite{Qiu-T4}.  Because the index $\sigma$ 
in Fig.~\ref{fig3} is transverse in light-cone gauge, 
the $\frac{1}{2}\gamma\cdot p'$ from the twist-2 antiquark
line eliminates the $\gamma\cdot p'$ term of the quark
propagator, and leaves the partonic contribution proportional
to $(x_2-x_1)/(x_2-x_1+i\epsilon)$.  The vanishing numerator 
factor $(x_2-x_1)$ cancels the potential
$1/(x_2-x_1)$ pole from the incoming gluon, and leaves a
single soft-gluon pole.  The derivative term in Eq.~(\ref{HammonA}) 
was obtained because the authors did not contract the incoming 
quark and antiquark lines with the $\frac{1}{2}\gamma\cdot p$
and $\frac{1}{2}\gamma\cdot p'$. For a detailed discussion on this
issue from the perspective of gauge vector ($n$) independence, see
Ref.~\cite{BM-dy-sts}. 

\section{asymmetry derived in a covariant gauge}

In deriving the single transverse-spin asymmetry in the Drell-Yan angular
distribution in the last section, we kept explicit electromagnetic 
current conservation.  In order to test the color gauge invariance of 
our result, we present a derivation of the same asymmetry in a color
covariant gauge in this section.

Similar to our calculation in light-cone gauge, we start with 
the general decomposition in Eq.~(\ref{DY-W-L}), and 
a complete set of partonic Feynman diagrams at the order of 
$O(g\alpha_{em}^2)$ in Fig.~\ref{fig3}.  Due to the change of 
the gauge, one difference is that 
the gluon field $A^\sigma(y_2^-)$ in 
the matrix element $T_q^\sigma(x_1,x_2;s_T)$ is no longer dominated
by its transverse components.  Instead, the leading contribution 
of the gluon field is from the component parallel to the direction 
of its momentum, $A^\sigma \propto p^\sigma$ \cite{LQS-DIS}.  
Therefore, we can get the leading contribution from the gluon field 
in a covariant gauge by the following replacement,
\begin{equation}
A^\sigma(y_2^-) \longrightarrow \frac{1}{p^+}\, 
  A^+(y_2^-)\, p^\sigma\, .
\label{A-cov}
\end{equation}
In order to convert the $A^+(y_2^-)$ to the corresponding field 
strength $F^{+\rho}(y_2^-)$, we need to give the gluon field a
small transverse momentum $k_T$, and then convert 
$k_T^\rho A^+(y_2^-)$ into $F^{+\rho}(y_2^-)$ \cite{QS-photon-sts}. 

With the gluon carrying a small transverse momentum $k_T$, 
the hadronic tensor $W^{\mu\nu}(Q)$ in Eq.~(\ref{DY-W-L}) 
has a slightly different expression,
\begin{equation}
W^{\mu\nu}(Q) = \sum_{q} e_q^2
   \int dx_1 dx_2 \int d^2k_T\, T_{q}(x_1,x_2,k_T;s_T) 
   \int dx'\, \bar{q}(x')\ 
   H^{\mu\nu}(x_1,x_2,k_T,x';Q)\, ,
\label{W-sts-fac-c}
\end{equation}
where $\sum_q$ runs over all quark and antiquark flavors.  
The hadronic matrix element $T_{q}(x_1,x_2,k_T;s_T)$ is 
defined as
\begin{eqnarray}
T_q(x_1,x_2,k;s_T) &=&
\int \frac{p^+dy_1^-}{2\pi}\, \frac{p^+dy_2^-}{2\pi} \,
     \frac{d^2y_T}{(2\pi)^2}\
     {\rm e}^{i x_1 p^+ y_1^-}\, {\rm e}^{i (x_2-x_1) p^+ y_2^-}\,
     {\rm e}^{i \vec{k}_T\cdot \vec{y}_T}
\nonumber \\
&\times &
\langle p,\vec{s}_T | \psibar_q(0,0_T) 
  \left(\frac{\gamma^+}{2p^+}\right)  
  \left[\frac{1}{p^+}A^+(y_2^-,y_T) \right] 
  \psi_q(y_1^-) | p,\vec{s}_T \rangle\, .
\label{T-q-def}
\end{eqnarray}
The partonic part $H^{\mu\nu}(x_1,x_2,k_T,x';Q)$ is again 
given by the four Feynman diagrams in Fig.~\ref{fig3}, except
the incoming gluon momentum is now $(x_2-x_1)p + k_T$, and  
the gluon polarization $\sigma$ is contracted with $p^\sigma$.

In order to extract the twist-3 contributions from these diagrams, 
we need to get one power of $k_T$ from the partonic part to 
convert the gluon field $A^+$ in the matrix element into 
corresponding gluon field strength \cite{QS-photon-sts}.  Expand 
the partonic part at $k_T=0$,
\begin{equation}
H^{\mu\nu} \approx 
H^{\mu\nu}(x_1,x_2,k_T=0,x';s_T)
+\frac{\partial}{\partial k_T^\rho}
    H^{\mu\nu}(x_1,x_2,k_T=0,x';s_T)\,
k_T^\rho
+O(k_T^2) .
\label{H-exp-c}
\end{equation}
The first term in the above expansion corresponds to the eikonal line 
contribution to the twist-2 quark distribution.  The second term 
in Eq.~(\ref{H-exp-c}) is responsible for the twist-3 contribution, 
while the remaining terms correspond to even higher twist contributions.

Substituting the second term of the above expansion into 
Eq.~(\ref{W-sts-fac-c}), and integrating over $\int d^2k_T$, we 
obtain the twist-3 contributions to the hadronic tensor in a 
covariant gauge,
\begin{equation}
W^{\mu\nu}(Q) = \sum_{q} e_q^2
   \int dx_1 dx_2 \, T_{F_q}^{(V)}(x_1,x_2) 
   \int dx'\, \bar{q}(x') \left[\left.
   \frac{i}{2\pi}\, 
   \epsilon^{\rho s_T n \bar{n}}\,
   \frac{\partial}{\partial k_T^\rho}
        H^{\mu\nu}\right|_{k_T=0} \right]\, .
\label{W-sts-c}
\end{equation}
where the twist-3 correlation function $T_{F_q}^{(V)}(x_1,x_2)$ 
is given in Eq.~(\ref{TFV-def}).  
Therefore, calculating the twist-3 short-distance contribution
is equivalent to extracting the terms linear in $k_T$ 
from the Feynman diagrams in Fig.~(\ref{fig3}) with incoming gluon
momentum $(x_2-x_1)p+k_T$.

As in the previous section, we calculate the short-distance partonic
contributions to the asymmetry in two steps.  We first evaluate
the contributions from the diagrams in Fig.~\ref{fig3}(a) and 
\ref{fig3}(b), and find 
\begin{eqnarray}
\left.
\frac{\partial}{\partial k_T^\rho}
   H^{\mu\nu}\right|_{k_T=0}(a+b) 
&=&  -g\,
\left( \frac{1}{x_2-x_1+i\epsilon} \right)
\left[ g^{\mu}_\rho \left(\frac{p}{x'}\right)^\nu
       \delta(Q^2-x_2x'S) \right.
\nonumber \\
&\ & {\hskip 1.3in} \left.
     + g^{\nu}_\rho \left(\frac{p}{x'}\right)^\mu
       \delta(Q^2-x_1x'S) \right].
\label{H-C-ab}
\end{eqnarray}
Again, the contributions from these two diagrams do not respect 
electromagnetic current conservation, and are exactly the same as 
what we calculated in the light-cone gauge.  Therefore, we need to 
find the other half of the contributions from the diagrams in 
Fig.~\ref{fig3}(c) and \ref{fig3}(d). 

In a covariant gauge, the gluon polarization index $\sigma$ in
Fig.~\ref{fig3} is contracted with $p^\sigma$.  For the two 
diagrams in Fig.~\ref{fig3}(c) and \ref{fig3}(d), the gluon 
interaction vertices are directly connected to the incoming quark
lines, which are to be contracted with $\frac{1}{2}\gamma\cdot p$.  
Naively, we would conclude that contributions from these two 
diagrams vanish because $p^\sigma[\gamma_\sigma\,\frac{1}{2}
\gamma\cdot p] \propto p^2 = 0$.  But, due to the vanishing 
denominator of the quark propagators of these two diagrams, we 
actually have a ``0/0'' situation for the contributions of these 
two diagrams.
  
\begin{figure}
\begin{center}
\epsfig{figure=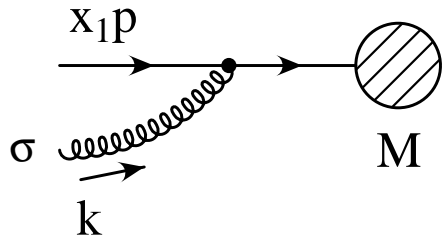,width=2.3in}
\end{center}
\caption{
A generic diagram for extracting the special 
quark propagator.
}
\label{fig5}
\end{figure}

Same as in the light-cone gauge, these two diagrams have not only 
twist-3 contributions but also long-distance twist-2 contributions. 
In order to extract the correct twist-3 contributions from these two 
diagrams in covariant gauge, consider a simple and generic diagram
in Fig.~\ref{fig5}, where the incoming quark of momentum $x_1p$ is 
on mass-shell, and the gluon's momentum 
$k \rightarrow (x_2-x_1)p + k_T$.  The contribution from this 
generic diagram can be written as
\begin{equation}
H \equiv {\rm Tr}\left[{M}(x_1p+k) \,
\frac{i\gamma\cdot(x_1p+k)}{(x_1p+k)^2+i\epsilon}\,
(-i\gamma_\sigma)\gamma\cdot(x_1p)\right]\, p^\sigma\, ,
\label{H-soft}
\end{equation}
where ${M}$ represents the blob in Fig.~\ref{fig5}.
Since we are interested in the twist-3 contribution in a covariant
gauge, we need only the terms linearly proportional to $k_T$,
\begin{eqnarray}
H \longrightarrow H^{(3)} 
&=& 
{\rm Tr}\left[{M}(x_2p)\gamma\cdot k_T\right]
\frac{x_1\, p^2}{(x_2-x_1)(2x_1p^2)+ i\epsilon}
\nonumber \\
&=&
{\rm Tr}\left[{M}(x_2p)\gamma\cdot k_T\right]
\frac{1}{2}\, \frac{1}{x_2-x_1+ i\epsilon}\, .
\label{H3-soft}
\end{eqnarray}
In deriving the first line in Eq.~(\ref{H3-soft}), we dropped 
a $k_T^2$ in the denominator and the $k_T$-dependence in ${M}$, 
because they correspond to twist-4 or higher contributions.

Following this approach, we derive the twist-3 contributions from
the two diagrams in Fig.~\ref{fig3}(c) and \ref{fig3}(d),
\begin{eqnarray}
\left.
\frac{\partial}{\partial k_T^\rho}
   H^{\mu\nu}\right|_{k_T=0}(c+d) 
&=& -g\,
\left( \frac{1}{x_2-x_1+i\epsilon} \right)
\left[ g^{\mu}_\rho \left(-\frac{p'}{x_2}\right)^\nu
       \delta(Q^2-x_2x'S) \right.
\nonumber \\
&\ & {\hskip 1.3in} \left.
     + g^{\nu}_\rho \left(-\frac{p'}{x_1}\right)^\mu
       \delta(Q^2-x_1x'S) \right].
\label{H-C-cd}
\end{eqnarray}
By adding all twist-3 contributions of the four diagrams in 
Fig.~\ref{fig3}, we obtain in a covariant gauge
\begin{eqnarray}
\left.
\frac{\partial}{\partial k_T^\rho}
   H^{\mu\nu}\right|_{k_T=0} 
&=& -g\,
\left( \frac{1}{x_2-x_1+i\epsilon} \right)
\left[ 
   g^{\mu}_\rho \left(\frac{p}{x'}-\frac{p'}{x_2}\right)^\nu
      \delta(Q^2-x_2x'S) \right.
\nonumber \\
&\ & {\hskip 1.3in} \left.
 + g^{\nu}_\rho \left(\frac{p}{x'}-\frac{p'}{x_1}\right)^\mu
      \delta(Q^2-x_1x'S) \right],
\label{H-C}
\end{eqnarray}
which clearly respects electromagnetic current conservation.  

By substituting Eq.~(\ref{H-C}) into Eq.~(\ref{W-sts-c}), and 
contracting with the leptonic tensor $L_{\mu\nu}$, 
we derive the same polarized Drell-Yan cross section in 
Eq.~(\ref{DY-sts-LO}) and the same asymmetry in 
Eq.~(\ref{BQ-AN-Full}) in a color covariant gauge.  Therefore, 
we can conclude that our result in Eq.~(\ref{BQ-AN-Full}) 
for the single transverse-spin asymmetry in the Drell-Yan lepton 
angular distribution not only respects electromagnetic gauge
invariance, but also is the same in both color light-cone 
and color covariant gauges.

\section{Numerical estimate and conclusions}

In order to estimate the numerical size of the single 
transverse-spin asymmetry in the Drell-Yan lepton angular distribution
in Eq.~(\ref{BQ-AN-Full}), we need to know the size and sign of the 
twist-3 quark-gluon correlation functions $T_{F_q}^{(V)}(x,x)$.

Fortunately, the same twist-3 quark-gluon correlation functions 
$T_{F_q}^{(V)}(x,x)$ appear in a number of other single 
transverse-spin asymmetries \cite{QS-photon-sts,QS-pi-sts}.  
In principle, self-consistency of the twist-3 correlation functions
in different physical observables is a test of QCD and its 
factorization theorems beyond the leading twist physics.  

In Ref.~\cite{QS-pi-sts}, single transverse-spin asymmetries in hadronic
pion production were calculated in perturbative QCD, in terms of the 
same generalized factorization theorem that was used in this paper. 
The same twist-3 quark-gluon correlation functions $T_{F_q}^{(V)}(x,x)$
dominate the leading contributions to the asymmetries.  Although the
existing data on pion asymmetries have relatively low pion transverse
momenta, the general features of the data are naturally explained by
the theoretical formula's \cite{QS-pi-sts}.  By comparing the theoretical 
calculations with the existing data on the difference between $\pi^+$ 
and $\pi^-$ asymmetries, it seems that the twist-3 correlation 
function $T_{F_u}^{(V)}(x,x)$ for valence up quark has an opposite 
sign to $T_{F_d}^{(V)}(x,x)$ for valence down quark.  In addition, 
a positive $T_{F_u}^{(V)}(x,x)$ is favored.

If we adopt the model introduced in Ref.~\cite{QS-pi-sts} for the 
twist-3 quark-gluon correlation functions,
\begin{equation}
T_{F_q}^{(V)}(x,x) \approx \kappa_q\, \lambda\, q(x)
\label{QS-model}
\end{equation}
with $\kappa_u=+1=-\kappa_d$, $\kappa_s=0$, 
and $\lambda \sim 100$~MeV,
we estimate the single transverse-spin asymmetry in the Drell-Yan
lepton angular distribution as follows. Taking the angle 
$\phi\sim 90$ degrees, we have the angular dependence,
\begin{equation} 
\left|\frac{\sin 2\theta\sin\phi}{1+\cos^2\theta}\right|\, 
\leq \, 0.7\, .
\label{AN-angle}
\end{equation}
Because the valence up quark is about twice the valence down quark, and 
the fractional charge square $e_u^2 =4/9$ is a factor of 4 
larger than $e_d^2=1/9$, we approximate
\begin{eqnarray}
\frac{\sum_q e_q^2 \int dx\, T_q(x,x)\, \bar{q}(Q^2/xS)}
     {\sum_q e_q^2 \int dx\, q(x)\, \bar{q}(Q^2/xS)}
&\sim &
-\, \lambda\, 
\frac{\int dx\, u(x)\, \bar{u}(Q^2/xS)}
     {\int dx\, \left[ u(x)\, \bar{u}(Q^2/xS) 
                     + \bar{u}(x)\, u(Q^2/xS) \right] }
\nonumber \\
&\sim &
-\, \frac{\lambda}{2}\, ,
\label{AN-pdf}
\end{eqnarray}
where the minus sign is due to the definition 
$T_q(x,x)=-\, T_{F_q}^{(V)}(x,x)$.  Because of the assumption, 
$k_{\bar{u}}=0$ \cite{QS-pi-sts}, we have only one term in the
numerator in Eq.~(\ref{AN-pdf}) in comparison with the two terms in
the denominator.  The 2 in the second line of Eq.~(\ref{AN-pdf}) is 
due to the fact that the two terms in the denominator
are equal for proton-proton collision.
Taking the maximum value from 
the angular dependence in Eq.~(\ref{AN-angle}), we estimate that 
the magnitude of the asymmetry in Eq.~(\ref{BQ-AN-Full}) 
can be as large as (for $\sqrt{4\pi\alpha_s}\sim 2$)
\begin{equation}
\left|A_N\right| \sim  0.7 \, \frac{\lambda}{Q}\, .
\label{AN-esti}
\end{equation}
For $\lambda\sim 100$~MeV, we have $A_N$ of about 3.5\% at $Q=2$~GeV, 
and 1.75\% at $Q=4$~GeV, where we chose the $Q$ values to be just
below and above the $J/\psi$ resonance. Considering that RHIC's
sensitivity to double transverse-spin asymmetries in Drell-Yan is at
the percent level \cite{Martin},  
these values for the single transverse-spin asymmetry might be
detectable.  

The numerical values for the single transverse-spin asymmetry in 
Drell-Yan lepton angular distribution are small, which is consistent
with the fact that the single transverse-spin asymmetries are a
twist-3 effect.  For any inclusive observables, like the Drell-Yan 
cross section discussed here, the asymmetries should be of the order
of $\Lambda_{\rm QCD}/Q$ with $Q$ being the energy exchange of the 
collisions.  However, the single transverse-spin asymmetries can be
large for observables at certain phase space where the twist-2
contributions are steeply falling, and the twist-3 contributions get 
a kinematic enhancement due to the derivative term \cite{QS-pi-sts}.

We would like to contrast our small asymmetry estimates with those of Ref.\
\cite{Boros}, which find large asymmetry values based on their model and its
parameters fitted to the pion production asymmetries. A future measurement of
the size of the asymmetry can therefore clearly distinguish between the model
predictions of \cite{Boros} (asymmetries between 20\% and 40\% for the whole
$x_F$ range) and the estimates presented here.  

In conclusion, we have calculated the single transverse-spin asymmetry
for the Drell-Yan lepton-pair's angular distribution in perturbative QCD.
We found that our result at leading order in the strong coupling
constant, given in Eq.~(\ref{BQ-AN-Full}), is consistent
with what was derived in Ref.~\cite{BMT-dy-sts}, but different from
what was found in Ref.~\cite{HTS-dy-sts}. 
We derived our result in both color light-cone and covariant gauges
while keeping explicit electromagnetic current conservation. 
With a model of the twist-3 correlation function, our calculated
asymmetry predicts both the sign and magnitude of the asymmetry.

\acknowledgments 

We would like to thank Oleg Teryaev for valuable discussions on this
subject.  
D.B.\ thanks the RIKEN-BNL Research Center, where this work was
started. At present, the research of D.B. has been made possible by a  
fellowship of the Royal Netherlands Academy of Arts and Sciences. 
J.Q. also thanks nuclear theory group at Brookhaven National 
Laboratory for support and hospitality while a part of this work 
was completed.    
The research of J.Q. at Iowa State was supported in part by 
the US Department of Energy under Grant No. DE-FG02-87ER40731.


\end{document}